\documentclass[12pt]{iopart}
\usepackage{graphicx}

\begin{document}
\def \ee {\varepsilon}
\thispagestyle{empty}
\title[Fast Track Communication]{
Thermal Casimir-Polder force between an atom and a dielectric
plate: thermodynamics and experiment
}

\author{
G.~L.~Klimchitskaya,${}^{1,2}$
U.~Mohideen${}^3$
and V.~M.~Mostepanenko${}^{1,4}$
}

\address{${}^1$Center of Theoretical Studies and Institute for Theoretical 
Physics, Leipzig University,
Postfach 100920, D-04009, Leipzig, Germany}

\address{$^2$North-West Technical University, Millionnaya St. 5,
St.Petersburg, 191065, Russia
}
\address{$^3$
Department of Physics and Astronomy, University of California, Riverside,
CA 92521, USA}

\address{$^4$
Noncommercial Partnership  ``Scientific Instruments'', 
Tverskaya St. 11, Moscow, 103905, Russia}

\begin{abstract}
The low-temperature behavior of the Casimir-Polder free energy
and entropy for an atom near a dielectric plate are found on the
basis of the Lifshitz theory.
The obtained results are shown to be thermodynamically
consistent if the dc
conductivity of the plate material is disregarded.
With inclusion of dc conductivity, both the standard Lifshitz
theory (for all dielectrics) and its generalization taking
into account screening effects (for a wide
range of dielectrics) violate the Nernst heat theorem.
The inclusion of the screening effects is
also shown to be 
inconsistent with experimental data of Casimir force measurements.
The physical reasons for this inconsistency are elucidated.
\end{abstract}

\section{Introduction}

Recently, considerable interest has been focused on the interaction
of atoms with metal and dielectric plates (walls) at separation
distances $a$ from about 100\,nm to a few micrometers where the
retardation effects of the electromagnetic field play an important
role. The relativistic description of the fluctuating interaction
of atoms with an ideal metal plate at zero temperature was pioneered
by Casimir and Polder \cite{1} who obtained the interaction
energy in the form $E(a)=-3\hbar c\alpha_0/(8\pi a^4)$, where
$\alpha_0\equiv\alpha(0)$ is the static atomic polarizability,
$\hbar$ and $c$ are the Planck constant and the velocity of light.
Lifshitz \cite{2} developed the general theory of dispersion forces
between two dielectric semispaces at a temperature $T$ in thermal
equilibrium with plane parallel boundary surfaces separated
by a gap $a$ of arbitrary width much larger than interatomic distances.
This theory describes material properties with a dielectric
permittivity that depends only on frequency. It includes both
nonrelativistic (when $a\ll\lambda_0$ where $\lambda_0$ is the
characteristic absorption wavelength of the semispace material) and
relativistic (when $a\gg\lambda_0$) limiting cases.
If the material of one of the plates is rarefied, the general
formula for the energy of the atom-wall interaction is obtained.
For an ideal metal plate at $T=0$ it leads to the Casimir-Polder
result. In the high temperature (large separation) limit
the interaction free energy of an atom with an ideal metal plate is
given by ${\cal F}(a,T)=-k_BT\alpha_0/(4a^3)$, where $k_B$ is
the Boltzmann constant. For a dielectric plate with the static
dielectric permittivity $\ee_0\equiv\ee(0)$, the free energy
acquires an additional factor $r_0=(\ee_0-1)/(\ee_0+1)$ which
goes to unity when $\ee_0\to\infty$.

In the last few years the Casimir-Polder and Lifshitz formulas
have been used for the interpretation of many experiments on quantum
reflection and Bose-Einstein condensation of ultra-cold atoms near
a variety of surfaces (see, e.g., \cite{3,5,6,8,10}
and references therein). The Lifshitz theory was also extensively
applied for the interpretation of measurements of the Casimir
force between two macrobodies made of metals and semiconductors
(see, e.g., \cite{11,11a,13,14,14a,15,15a,16,17,18}). All these experiments
have attracted widespread attention from diverse fields ranging from
condensed matter physics and nanotechnology to atomic physics,
elementary particle physics, gravitation and cosmology.
However, the application of the Lifshitz theory to the real material
bodies used in experiments has resulted in a puzzle. It was found
that for metallic test bodies with perfect crystal lattices
the calculational results are in
contradiction with thermodynamics
\cite{19,20} and experimental data \cite{13,14,14a} if the
relaxation processes of conduction electrons are included into
the model of the dielectric response used in the Lifshitz theory.
For semiconductor and dielectric materials whose conductivity
goes to zero with vanishing temperature, the calculational
results using the Lifshitz theory were also shown to be in
contradiction with thermodynamics \cite{21,21a,22} and
experiment \cite{16,17,23} when the dc conductivity of a dielectric or
a high-resistivity semiconductor plate is included.

An interesting attempt to find the physical explanation for some
of these puzzling results is undertaken in \cite{24}.
It takes into account the Debye screening of the electrostatic field
by free charge carriers in the conductor leading to a modified
reflection coefficient for the transverse magnetic (TM) mode
at zero frequency in comparison to the Lifshitz theory 
(the above mentioned puzzles arise
only in the zero-frequency term of the Lifshitz formula).
The reflection coefficient for the transverse electric
(TE) mode at zero frequency for dielectric materials is equal
to zero regardless of the dc conductivity.

In this paper we find the low-temperature behavior of the entropy of the
Casimir-Polder atom-plate interaction both in the framework of the
commonly accepted Lifshitz theory and its generalization taking
into account the screening effects.
We demonstrate that if the dc
conductivity of dielectrics is disregarded 
 the Casimir-Polder entropy goes to zero
when the temperature vanishes, i.e., the Nernst heat theorem is
satisfied. Thus, we show that in this case the standard Lifshitz theory
of atom-wall interaction is in agreement with thermodynamics. 
If the dc conductivity is included, the standard
Lifshitz theory violates Nernst's theorem. The generalization
\cite{24} is in agreement with Nernst's theorem for die\-lec\-t\-rics
whose charge carrier density vanishes when $T$ goes to zero.
However, for dielectrics whose charge carrier density is
temperature-independent (for such materials conductivity goes
to zero with $T$ not due to the vanishing $n$ but due to the
vanishing mobility of the charge carriers) the generalization of
the Lifshitz theory taking into account the screening
effects is shown to violate the Nernst theorem.  
 Finally we demonstrate that
the suggested generalization of TM reflection coefficient at zero
frequency is inconsistent with the measurement data of the
difference Casimir force between a metal sphere and a semiconductor
plate \cite{16,17}.

\section{Low-temperature behavior of the Casimir-Polder entropy}

We start with the standard Lifshitz formula for the free energy
of an atom-plate interaction \cite{2,5,6}
\begin{equation}
{\cal F}(a,T)=\frac{k_BT}{8a^3}\sum_{l=0}^{\infty}
{\vphantom{\sum}}^{\prime}\Phi_{A}(\zeta_l),
\label{eq1}
\end{equation}
\noindent
where $\zeta_l=\xi_l/\omega_c$, $\xi_l=2\pi k_BTl/\hbar$ are the
Matsubara frequencies, $\omega_c=c/(2a)$, prime indicates that the
$l=0$ term has to be multiplied by 1/2, and
\begin{equation}
\hspace*{-1.3cm}
\Phi_{A}(x)=-\alpha({\rm i}\omega_cx)\int_{x}^{\infty}dy{\rm e}^{-y}
\left\{2y^2r_{\rm TM}({\rm i}x,y)
-x^2\left[r_{\rm TM}({\rm i}x,y)+r_{\rm TE}({\rm i}x,y)
\right]\right\}.
\label{eq2}
\end{equation}
\noindent
The reflection coefficients are defined through the frequency-dependent
permittivity
$\ee\equiv\ee({\rm i}\omega_cx)$
\begin{equation}
\hspace*{-1.3cm}
r_{\rm TM}({\rm i}x,y)=\frac{\ee y-\sqrt{y^2+
x^2(\ee-1)}}{\ee y+\sqrt{y^2+x^2(\ee-1)}},
\qquad
r_{\rm TE}({\rm i}x,y)=\frac{y-\sqrt{y^2+
x^2(\ee-1)}}{y+\sqrt{y^2+x^2(\ee-1)}}.
\label{eq3}
\end{equation}
\noindent
The atomic dynamic polarizability
can be represented with sufficient precision
using the single-oscillator model \cite{5}
\begin{equation}
\alpha({\rm i}\omega_c\zeta_l)=\frac{\alpha_0}{1+\beta_A^2\zeta_l^2}
\label{eq4}
\end{equation}
\noindent
with a dimensionless constant $\beta_A$.

Using the Abel-Plana formula \cite{27}, the free energy (\ref{eq1})
can be equivalently represented in the form
\begin{equation}
{\cal F}(a,T)=E(a)+{\rm i}\frac{k_BT}{8a^3}\int_{0}^{\infty}
\!\!dt\frac{\Phi_A({\rm i}\tau t)-
\Phi_A(-{\rm i}\tau t)}{{\rm e}^{2\pi t}-1},
\label{eq5}
\end{equation}
\noindent
where $E(a)$ is the Casimir-Polder energy at zero temperature,
$\tau=2\pi T/T_{\rm eff}$ and the effective temperature is defined
from $k_BT_{\rm eff}=\hbar\omega_c$.

We are interested in the primary contribution to the low-temperature
asymptotic behavior of the Casimir-Polder free energy (\ref{eq1})
in the case $\ee_0<\infty$ (i.e., with dc conductivity
disregarded). For this purpose, as shown in \cite{21,21a,22}, it is
sufficient to restrict ourselves to a frequency-independent permittivity
$\ee({\rm i}\xi_l)=\ee_0$. By expanding $\Phi_A(x)$ in (\ref{eq2}) in
powers of $x$ and using (\ref{eq3}) and (\ref{eq4}),
one obtains
\begin{equation}
\Phi_A(x)=-\alpha_0\left[4r_0+r_0\left(4\beta_A^2-2
\frac{\ee_0}{\ee_0+1}-1\right)x^2+C_D(\ee_0)x^3\right],
\label{eq6}
\end{equation}
\noindent
where
the terms in higher powers are omitted and the following
notation is introduced
\begin{eqnarray}
&&
C_D(\ee_0)=r_0\frac{7\ee_0+1}{3(\ee_0+1)}+
\frac{(\sqrt{\ee_0}-1)\left[(3\ee_0^2+1)(2\sqrt{\ee_0}+1)+
3\ee_0(\sqrt{\ee_0}-1)\right]}{3(\sqrt{\ee_0}+1)(\ee_0+1)^2}
\nonumber \\
&&~~~
+\frac{2\ee_0^2}{(\ee_0+1)^{5/2}}\left(
{\rm Artanh}\sqrt{\frac{\ee_0}{\ee_0+1}}-
{\rm Arcoth}\sqrt{\ee_0+1}\right).
\label{eq7}
\end{eqnarray}
\noindent
In the limiting case $\ee_0\to 1$ we have $C_D(\ee_0)\to 0$ as
expected. The typical values of this coefficient are
$C_D(\ee_0)=0.585$ and 7.60 for $\ee_0=1.5$ and 16, respectively.
For the commonly used dielectrics such as SiO${}_2$ with $\ee_0=3.81$
and Si with $\ee_0=11.67$, from (\ref{eq7}) we get
$C_D(\ee_0)=2.70$ and 6.33, respectively.

As a result, from (\ref{eq6}) we obtain
\begin{equation}
\Phi_A({\rm i}\tau t)-\Phi_A(-{\rm i}\tau t)=2\tau^3t^3
\alpha_0C_D(\ee_0).
\label{eq8}
\end{equation}
\noindent
Then from (\ref{eq5}) the Casimir-Polder free energy is given by
\begin{equation}
{\cal F}(a,T)=E(a)-\frac{\hbar c\pi^3}{240a^4}\alpha_0C_D(\ee_0)
\left(\frac{T}{T_{\rm eff}}\right)^4
\label{eq9}
\end{equation}
\noindent
and the Casimir-Polder entropy by
\begin{equation}
S(a,T)=-\frac{\partial{\cal F}(a,T)}{\partial T}=
\frac{\pi^3k_B}{30a^3}\alpha_0C_D(\ee_0)
\left(\frac{T}{T_{\rm eff}}\right)^3.
\label{eq10}
\end{equation}
\noindent
As can be seen from (\ref{eq10}), the entropy goes to zero when $T$
vanishes in accordance with the Nernst heat theorem. Thus, the
Lifshitz theory of the atom-plate interaction is thermodynamically
consistent if the dc conductivity of dielectric plate is
disregarded.

In electrodynamics the inclusion of the dc conductivity is
equivalent to the replacement of $\ee(\omega)$ with
\begin{equation}
\tilde{\ee}(\omega,T)={\ee}(\omega)+
\frac{4\pi{\rm i}\sigma_0(T)}{\omega}.
\label{eq11}
\end{equation}
\noindent
In the Lifshitz theory this leads to only negligible additions to
all the terms at $\omega={\rm i}\xi_l$ with $l\geq 1$
in the free energy and entropy.  These additions
exponentially decay to zero with vanishing temperature
\cite{21,21a,22}. However, the term with $l=0$ is
modified because according to (\ref{eq3})
$r_{\rm TM}(0,y)=r_0$ is replaced with
$\tilde{r}_{\rm TM}(0,y)=1$. As a result, with dc conductivity
included the free energy of the atom-plate interaction at low
temperature is given by
\begin{equation}
\tilde{\cal F}(a,T)={\cal F}(a,T)-\frac{k_BT}{4a^3}(1-r_0)\alpha_0,
\label{eq12}
\end{equation}
\noindent
where ${\cal F}(a,T)$ is defined in (\ref{eq9}).
{}From (\ref{eq12}) one immediately arrives at the violation of the
Nernst heat theorem
\begin{equation}
\tilde{S}(a,0)=\frac{k_B\alpha_0}{4a^3}(1-r_0)>0.
\label{eq13}
\end{equation}

\section{Attempt to include the screening effects}

Now we apply the above thermodynamic test to the generalization of
the Lifshitz theory taking into account the screening effects. 
As is well known,
such effects arise when a static electric field penetrates
into a medium with a nonzero concentration of charge carriers.
In this case, the reflection coefficient $r_{\rm TM}(0,y)$,
as given in (\ref{eq3}) (the standard Lifshitz theory),
is modified to
\begin{equation}
r_{\rm TM}^{\rm mod}(0,y)=\frac{\ee_0\sqrt{4a^2\kappa^2+y^2}-
y}{\ee_0\sqrt{4a^2\kappa^2+y^2}+y},
\label{eq14}
\end{equation}
\noindent
where $\kappa^2=4\pi e^2n/(\ee_0k_BT)$ and $n$ is the total density of
free charge carriers, 
while all the coefficients
$r_{\rm TM,TE}({\rm i}\zeta_l,y)$ with $l\geq 1$ and also
 $r_{\rm TE}(0,y)=0$ remain unchanged \cite{24}.
Here, $\kappa$ is
connected with the so-called {\it Debye radius}, $\kappa=1/R_D$.
When $n=0$, (\ref{eq14})
leads to the same result as (\ref{eq3}). For $n\to\infty$,
at fixed $T\neq 0$, $r_{\rm TM}^{\rm mod}(0,y)=1$, as
in the case of the standard Lifshitz theory when the dc conductivity
is included.

The calculation of the free energy at low temperature with the
modified reflection coefficient (\ref{eq14}) results in
\begin{equation}
{\cal F}^{\rm mod}(a,T)={\cal F}(a,T)-\frac{k_BT\alpha_0}{8a^3}
\int_{0}^{\infty}\!r_{\rm TM}^{\rm mod}(0,y){\rm e}^{-y}y^2dy
+\frac{k_BT\alpha_0}{4a^3}r_0,
\label{eq15}
\end{equation}
\noindent
where ${\cal F}(a,T)$ is defined in (\ref{eq9}).
The respective Casimir-Polder entropy is given by
\begin{eqnarray}
&&
S^{\rm mod}(a,T)=S(a,T)+\frac{k_B\alpha_0}{4a^3}\left[\frac{1}{2}
\int_{0}^{\infty}\!r_{\rm TM}^{\rm mod}(0,y){\rm e}^{-y}y^2dy
-r_0\right]
\nonumber \\
&&~~~~~~~~
+\frac{k_BT\alpha_0}{8a^3}
\int_{0}^{\infty}
\frac{\partial r_{\rm TM}^{\rm mod}(0,y)}{\partial T}
{\rm e}^{-y}y^2dy,
\label{eq16}
\end{eqnarray}
\noindent
with $S(a,T)$ defined in (\ref{eq10}).
It is easily seen that the last term on the right-hand side of
(\ref{eq16}) goes to zero when temperature vanishes, regardless
of the temperature dependence of $n$. The second term on
the right-hand side of (\ref{eq16}) is more involved.
If $n(T)$ exponentially decays to zero with vanishing temperature
(as is true for insulators and intrinsic semiconductors),
then so does $\kappa(T)$. As a result,
$r_{\rm TM}^{\rm mod}(0,y)\to r_0$ and the entropy
$S^{\rm mod}(a,0)$ is equal to zero, in accordance with  the Nernst
theorem. However, if $n$ does not go to zero when $T$ goes to zero
(this is true, for instance, for dielectric materials, such as
semiconductors doped below critical
and solids with ionic conductivity), $\kappa\to\infty$ with
vanishing temperature and $r_{\rm TM}^{\rm mod}(0,y)\to 1$ when
$T\to 0$. In this case we obtain from (\ref{eq16})
\begin{equation}
S^{\rm mod}(a,0)=\tilde{S}(a,0)=\frac{k_B\alpha_0}{4a^3}(1-r_0)>0,
\label{eq17}
\end{equation}
i.e., the proposed generalization violates the Nernst heat theorem in
the same way as the standard Lifshitz theory with included dc conductivity
[compare with (\ref{eq13})]. In fact, conductivity
$\sigma_0(T)=n|e|\mu$, where $\mu$ is a mobility of charge carriers
\cite{28}. Although $\sigma_0(T)$ goes to zero exponentially
fast for all dielectrics when $T$ goes to zero, for most of them
this happens due to the vanishing mobility. For instance, the
conductivity of SiO${}_2$ (used in the calculations \cite{24} and,
as the plate material, in the experiment \cite{10})
is ionic in nature and is determined by the concentration of
impurities (alkali ions) which are always present as trace constituents.
According to our result, for this material the generalization
of the Lifshitz theory proposed in \cite{24} violates the Nernst
theorem. 

The existence of dielectric materials for which $n$ does not go to zero
but $\mu$ does go to zero when $T$ vanishes demonstrates that the
reflection coefficient (\ref{eq14}) is ambiguous. In reality, for such
materials $r_{\rm TM}^{\rm mod}(0,y)\to 1$ when $T$ and 
$\mu$ simultaneously
vanish. This is because $\kappa\to\infty$ when $T\to 0$ in
disagreement with physical intuition that there should be no screening
at zero mobility. In the conclusions, this ambiguity is
linked to the break in continuity of the more general reflection coefficients
at the point $\omega=0$, $T=0$.

The reflection coefficient (\ref{eq14}) can be formally obtained
from the standard Lifshitz theory with dissimilar permittivities
$\ee_x=\ee_y$ and $\ee_z$ commonly used for the description of
uniaxial crystals \cite{28a,28b}. In this case the transverse magnetic
reflection coefficient at zero frequency is given by
\begin{equation}
r_{\rm TM}^{\rm uni}(0,y)=
\frac{\sqrt{\ee_{0x}\ee_{0z}}-1}{\sqrt{\ee_{0x}\ee_{0z}}+1}.
\label{eq17a}
\end{equation}
\noindent
Then (\ref{eq14}) follows from (\ref{eq17a}) if one puts
\begin{equation}
\ee_{0x}=\ee_0,\qquad
\ee_{0z}=\ee_0\left(1+\frac{4a^2\kappa^2}{y^2}\right).
\label{eq17b}
\end{equation}
\noindent
However, an important difference of (\ref{eq17a}), (\ref{eq17b}) from
permittivities used in the case of uniaxial crystals is that (\ref{eq17b})
 depends on the wave vector. This is in fact an extension of
the Lifshitz theory with the inclusion of spatial dispersion.
Such an inclusion is
controversial and has been debated
for long in the literature (see the negative
conclusions regarding such an inclusion in \cite{29}
and the recent discussion \cite{30}).

The question of whether there is a possibility to compare
the theoretical predictions of \cite{24} with experimental
data should be considered.
This could be done with regard to the experiments
on measuring the Casimir-Polder interaction between an atom and
a SiO${}_2$ plate \cite{10} and the Casimir interaction between an
Au-coated sphere and a Si plate \cite{15,15a,15b}. In both cases the TE
reflection coefficient at zero frequency does not contribute to
the theoretical result. The experiment \cite{10} was successfully
used \cite{23} to demonstrate that the inclusion of the dc conductivity
of SiO${}_2$ in the standard Lifshitz theory is inconsistent with the data.
This is in support of our result in section 2 that such an inclusion
leads to a contradiction between the Lifshitz theory and thermodynamics.
In this connection the phenomenological prescription was formulated 
\cite{35a} that for dielectric materials the conductivity arising
at nonzero temperature should be disregarded in the calculation of
the Casimir force using the Lifshitz theory.

Recently one more experiment has been performed on measuring the
difference of the Casimir forces between an Au sphere and B doped
p-type Si plate illuminated with laser pulses \cite{16,17}. In the
absence of laser pulse the concentration of charge carriers in a Si
plate was $\tilde{n}=5\times 10^{14}\,\mbox{cm}^{-3}$ and in the
presence of pulse  $n_1=2.1\times 10^{19}\,\mbox{cm}^{-3}$ or
$n_2=1.4\times 10^{19}\,\mbox{cm}^{-3}$ for two different
absorbed powers $P_{w1}=9.3\,$mW and $P_{w2}=4.7\,$mW.
The directly measured quantity was $\Delta F(a)=F^{L}(a)-F(a)$
where $F^{L}$ and $F$ are the Casimir forces in the presence and in
the absence of laser light on the plate, respectively. The experimental
data were compared with the Lifshitz theory with neglected or included
dc conductivity of the high-resistivity Si in  the dark phase.
In the latter case the theoretical model was found to be inconsistent
with the data.
Here, we compare the measurement data of this experiment (shown as crosses
in Fig.\ 1 with experimental errors in force measurement determined
at 70\% confidence level) with computations on the basis of the
standard Lifshitz theory with the dc conductivity neglected in the dark
phase (solid lines) and on the basis of its generalization taking
into account the screening effects \cite{24} (dashed lines).
Figs.\ 1(a) and 1(b) are related to the absorbed powers  $P_{w1},\>P_{w2}$,
respectively. The experiment was performed in a high vacuum at $T=300\,$K
(see \cite{17} for details). The solid lines were computed using the standard
Lifshitz formula for the Casimir force with Si in the dark phase described
as dielectric, i.e., by $\ee({\rm i}\xi_l)$ with $\ee(0)=\ee_0<\infty$.
In the presence of light the dielectric permittivity
\begin{equation}
\ee^{L}({\rm i}\xi_l)=\ee({\rm i}\xi_l)+
\frac{\omega_{p(e)}^2}{\xi_l^2}+\frac{\omega_{p(p)}^2}{\xi_l^2}
\label{eq18}
\end{equation}
\noindent
has been used with the values of plasma frequencies for electrons and
holes determined in \cite{17} for different absorbed powers.
Almost the same results are obtained if the Drude description of
charge carriers is used in the presence of light
\begin{equation}
\tilde{\ee}({\rm i}\xi_l)=\ee({\rm i}\xi_l)+
\frac{\omega_{p(e)}^2}{\xi_l[\xi_l+\gamma^{(e)}]}+
\frac{\omega_{p(p)}^2}{\xi_l[\xi_l+\gamma^{(p)}]}
\label{eq19}
\end{equation}
\noindent
(see \cite{17} for the values of all parameters at different absorbed powers).
The dashed lines are obtained using Eq.\ (\ref{eq14}) for the zero frequency
TM reflection coefficient with different concentrations of charge
carriers $n=\tilde{n}$ in the dark phase and $n=2n_1$ or $2n_2$ in the
presence of light. At all nonzero Matsubara frequencies, in accordance
with \cite{24}, the standard terms of the Lifshitz formula were used.
The gold coated sphere was described by the commonly used dielectric
permittivity along the imaginary frequency axis (see, e.g., \cite{14a,17,18}).
We have verified that for Au the use of expression (\ref{eq14}) instead
of the standard zero-frequency term, as given by the Lifshitz theory, leads
to numerically the same results up to five significant figures.

\begin{figure*}[t]
\vspace*{-5.4cm}
\includegraphics{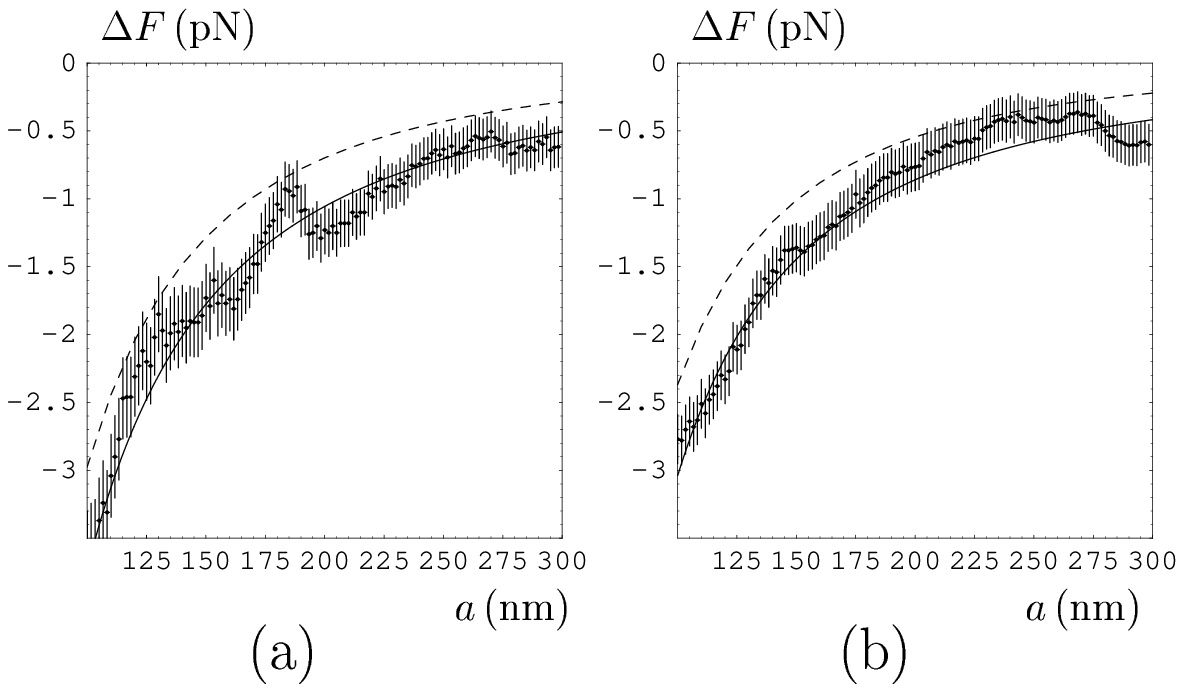}
\vspace*{-17.5cm}
\caption{
Difference of the Casimir forces between an Au-coated sphere
and a Si plate in the presence and in the absence of laser light on
the plate versus separation (a) for the absorbed power of 9.3\,mW and (b)
4.7\,mW.  The experimental data are shown as crosses.
Solid and dashed lines indicate the theoretical results calculated using
the standard Lifshitz theory with the dc conductivity of Si in the
dark phase neglected and the generalization of this theory \cite{24},
respectively.
}
\end{figure*}
As is seen in figure~1(a,b), the experimental data are consistent with the
theoretical results computed on the basis of the standard Lifshitz theory
with the dc conductivity of dielectric Si neglected in the dark phase
(the solid lines). The theoretical results computed on the basis of the
generalized Lifshitz theory \cite{24} with the modified TM reflection
coefficient at zero frequency are excluded by data at a 70\% confidence
level. The same conclusion follows from the third data set obtained
in Ref.\ \cite{17} at $P_{w3}=8.5\,$mW absorbed power.

\section{Conclusions and discussion}

To conclude, we have found the low-temperature behavior of the
Casimir-Polder free energy and entropy in atom-plate configuration
on the basis of the Lifshitz theory. For a dielectric plate with the
dc conductivity of the plate material neglected, the Lifshitz theory was
shown to be thermodynamically consistent. With the dc conductivity included,
the standard Lifshitz theory of atom-plate interaction violates the
Nernst heat theorem. The thermodynamic test was also applied to the
recent generalization of the Lifshitz theory taking into account the
screening effects. The proposed generalization
is shown to be in violation of the Nernst theorem for a wide range
of dielectric materials including doped semiconductors with doping
concentration below critical and ionic conductors. This generalization
is also inconsistent with the measurements of the difference Casimir force
between a metal sphere and Si plate illuminated with laser pulses.
Recently, the generalization of the Lifshitz theory \cite{24}
was extended using the Boltzmann transport equation \cite{31}.
 This approach, in addition to the standard
drift current $\mbox{\boldmath$j$}$, takes into account the
diffusion current $eD\nabla n$, where $D$ is the diffusion constant.
In the quasi-static limit $\omega\to 0$
the reflection coefficient (\ref{eq14}) was reproduced.
However, if one takes the limit $T\to 0$ first, keeping 
$\omega={\rm const}\neq 0$, the standard Fresnel reflection coefficients
(\ref{eq3}) on the dielectric surface with no screening are obtained.
This property is preserved in the subsequent limiting
transition $\omega\to 0$.
The calculation of the difference Casimir force in the experimental
configuration \cite{16,17} using the reflection coefficients \cite{31}
defined at all Matsubara frequencies leads to precisely the same results,
as shown by the dashed lines in Figure~1(a,b). Thus, the theoretical
approach \cite{31} is also inconsistent with experiment.
This is in line with the above phenomenological prescription stating
that conductivity arising in dielectric materials at $T\neq 0$
should be disregarded.

One may attempt to explain
the physical reason of why the Lifshitz theory does not allow the
inclusion of real conductivity processes as follows.
Lifshitz derived his famous formulas under the condition of thermal
equilibrium. This means that not only $T={\rm const}$, but also all
irreversible processes connected with the dissipation of energy
into heat have already been terminated \cite{38,39}. 
The Drude-like dielectric
function (\ref{eq11}) is derived from the Maxwell equations with a
real drift current of conduction electrons 
$\mbox{\boldmath$j$}=\sigma_0\mbox{\boldmath$E$}$ initiated by the
external electric field $\mbox{\boldmath$E$}$ \cite{26}.
The drift current is an irreversible process which
brings a system out of thermal equilibrium.
The characteristic of this irreversible process, the Drude type of 
dielectric function, can be also derived within the
Kubo formalism \cite{40} using the correlation function of the electric field 
averaged over the thermodynamic state functions which are assumed to have 
the same values as they would have in an equilibrium situation.
This allows a theoretical description of irreversible processes under
the assumption that there is local equilibrium. 
The real current leads to Joule's heating of the Casimir plates
(Ohmic losses) \cite{40a}.
To preserve the temperature constant, one should admit that there
exists an unidirectional
flux of heat from the medium to the heat reservoir \cite{40b}. 
Such interactions between
a system and a heat reservoir are prohibited by the definition of
 thermal equilibrium. 
Although the screening and diffusion effects really occur in an
external electric field, they are also related to physical situations 
out of thermal equilibrium. The reason is that the diffusion
current is determined by a nonzero gradient of charge carrier density,
whereas for homogeneous systems in thermal equilibrium the charge
carrier density must be homogeneous.

Our conclusion is that the standard Lifshitz theory of both atom-wall 
and wall-wall interactions is in a very good agreement with both
thermodynamics and all available experiments. Violations of the
Nernst theorem and contradictions with the experimental data arise
when the theory is applied to physical phenomena involving electric 
current, be it of drift or diffusion type.  Such an application 
seems to be in 
violation of thermal equilibrium which is the basic applicability condition
of the Lifshitz theory. 

\section*{Acknowledgments}
This work was supported by the NSF Grant No PHY0653657
(computation of the Casimir-Polder force), and the DOE
Grant No DE-FG02-04ER46131 (measurements of difference
Casimir force between Si and Au).
G.L.K.\ and V.M.M.\ were also supported by the
DFG Grant No.~436\,RUS\,113/789/0--4.

\end{document}